\documentclass[
  aps,
  prx,
  twocolumn,
  reprint,
  amsfonts,
  amssymb,
  amsmath,
]{revtex4-2}

\usepackage[utf8]{inputenc}
\usepackage{bm}  
\usepackage{siunitx}
\usepackage{booktabs}
\usepackage{graphicx}
\usepackage[svgnames]{xcolor}

\usepackage[hidelinks]{hyperref}

\newcommand*{\vb}[1]{\bm{#1}}  
\newcommand*{\dd}{\mathrm{d}}  
\newcommand*{\pdv}[2]{\frac{\partial #1}{\partial #2}}
\newcommand*{\laplacian}{\nabla^2}
\newcommand*{\gradient}{\vb{\nabla}}

\newcommand*{\Rey}{\textit{Re}}
\newcommand*{\ReLambda}{\Rey_\lambda}
\newcommand*{\TaylorScale}{\lambda_{\text{T}}}

\newcommand*{\Pcirc}{\mathcal{P}_r}  
\newcommand*{\Prob}{\mathbb{P}}

\newcommand*{\resampling}{\beta}

\newcommand*{\vvec}{\vb{v}}
\newcommand*{\kvec}{\vb{k}}
\newcommand*{\xvec}{\vb{x}}
\newcommand*{\fvec}{\vb{f}}
\newcommand*{\vortvec}{\vb{\omega}}

\newcommand*{\rvec}{\vb{r}}
\newcommand*{\Cloop}{\mathcal{C}}
\newcommand*{\CircA}{\Gamma_{\!A}}
\newcommand*{\CircR}{\Gamma_{\!r}}
\newcommand*{\lambdaKolm}{\lambda^{\text{K41}}}

\newcommand*{\mean}[1]{\langle #1 \rangle}
\newcommand*{\Lint}{\ell_\text{I}}  
\newcommand*{\Vnl}{V_{\mathrm{I}}}

\newcommand*{\qfor}{\quad \text{ for }}

\begin{document}

\title{Intermittency of velocity circulation in quantum turbulence}

\author{Nicol\'as P. M\"uller}
\author{Juan Ignacio Polanco}
\author{Giorgio Krstulovic}
\affiliation{%
  Université Côte d'Azur, Observatoire de la Côte d'Azur, CNRS,
  Laboratoire Lagrange, Boulevard de l'Observatoire CS 34229 - F 06304 NICE Cedex 4, France
}
\date{\today}

\begin{abstract}
  The velocity circulation, a measure of the rotation of a fluid within a closed path, is a fundamental observable in classical and quantum flows.
  It is indeed a Lagrangian invariant in inviscid classical fluids.
  In quantum flows, circulation is quantized, taking discrete values that are directly related to the number and the orientation of thin vortex filaments enclosed by the path.
  By varying the size of such closed loop, the circulation provides a measure of the dependence of the flow structure on the considered scale.
  Here we consider the scale dependence of circulation statistics in quantum turbulence, using high resolution direct numerical simulations of a generalized Gross--Pitaevskii model.
  Results are compared to the circulation statistics obtained from simulations of the incompressible Navier--Stokes equations.
  When the integration path is smaller than the mean inter-vortex distance, the statistics of circulation in quantum turbulence displays extreme intermittent behavior due to the quantization of circulation, in stark contrast with the viscous scales of classical flows.
  In contrast, at larger scales, circulation moments display striking similarities with the statistics probed in the inertial range of classical turbulence.
  This includes the emergence of the power law scalings predicted from Kolmogorov's 1941 theory, as well as intermittency deviations that closely follow the recently proposed bifractal model for circulation moments in classical flows.
  To date, this is the most convincing evidence of intermittency in the large scales of quantum turbulence.
  Moreover, our results strongly reinforce the resemblance between classical and quantum turbulence, highlighting the universality of inertial range dynamics, including intermittency, across these two a priori very different systems.
  This work paves the way for an interpretation of inertial range dynamics in terms of the polarization and spatial arrangement of vortex filaments.
\end{abstract}

\maketitle

\section{Introduction}

The motion of vortices in fluid flows, including rivers, tornadoes, and the outer atmosphere of planets like Jupiter, has fascinated observers for centuries. Vortices are a defining feature of turbulent flows, and their dynamics and their mutual interaction are the source of very rich physics.
One notable example of such interaction is the reconnection between vortex filaments~\cite{Kida1994}, the process by which a pair of vortices may induce a change of topology following their mutual collision.
In inviscid classical fluids, Helmholtz's theorems~\cite{Helmholtz1858} imply that a vortex tube preserves its identity over time, thus disallowing reconnections. An extension of this result is Kelvin's theorem~\cite{Thomson1868}, which states that the velocity circulation about a closed loop moving with the flow is conserved in time.
The velocity circulation about a closed loop $\Cloop$ enclosing an area $A$, defined from the fluid velocity $\vvec$ by
\begin{equation}
  \label{eq:circulation_intro}
  \CircA(\Cloop; \vvec) = \oint_\Cloop \vvec \cdot \dd\rvec,
\end{equation}
is directly related to the vorticity flux across the loop via Stokes' theorem, and thus to the topology and the dynamics of vortex filaments.
In non-ideal classical flows, one effect of viscous dissipation is to smooth-out the interface between vortices and the surrounding fluid.
As a result, vortex reconnections become possible, and the circulation is no longer conserved about advected loops.

Superfluids, such as very low temperature liquid helium, have the astonishing property of being free of viscous dissipation.
This is closely related to Bose--Einstein condensation, and is a clear manifestation of quantum physics at macroscopic scales.
As a result, superfluids can be effectively described by a macroscopic wave function.
This description supports the emergence of \emph{quantum vortices}, topological defects where the wave function vanishes, which in three-dimensional space take the form of thin filaments.
Moreover, the velocity circulation about such vortices is quantized in units of the quantum of circulation $\kappa = h / m$, where $h$ is Planck's constant, and $m$ is the mass of the bosons constituting the superfluid~\cite{Barenghi2014}.

Despite the absence of viscosity, it is now well known that vortices in superfluids can reconnect.
This possibility was initially suggested by Feynman~\cite{Feynman1955}, and was first verified numerically in the frame of the Gross--Pitaevskii (GP) model~\cite{Koplik1993}.
Quantum vortex reconnections were later visualized experimentally in liquid helium~\cite{Bewley2008} as well as in trapped Bose--Einstein condensates~\cite{Serafini2017}.
Vortex reconnections are considered to be an essential mechanism for sustaining the whole turbulent process~\cite{Schwarz1988,Donnelly1993,Nore1997a}.

\begin{figure*}[t]
  \centering
  \includegraphics[width=\textwidth]{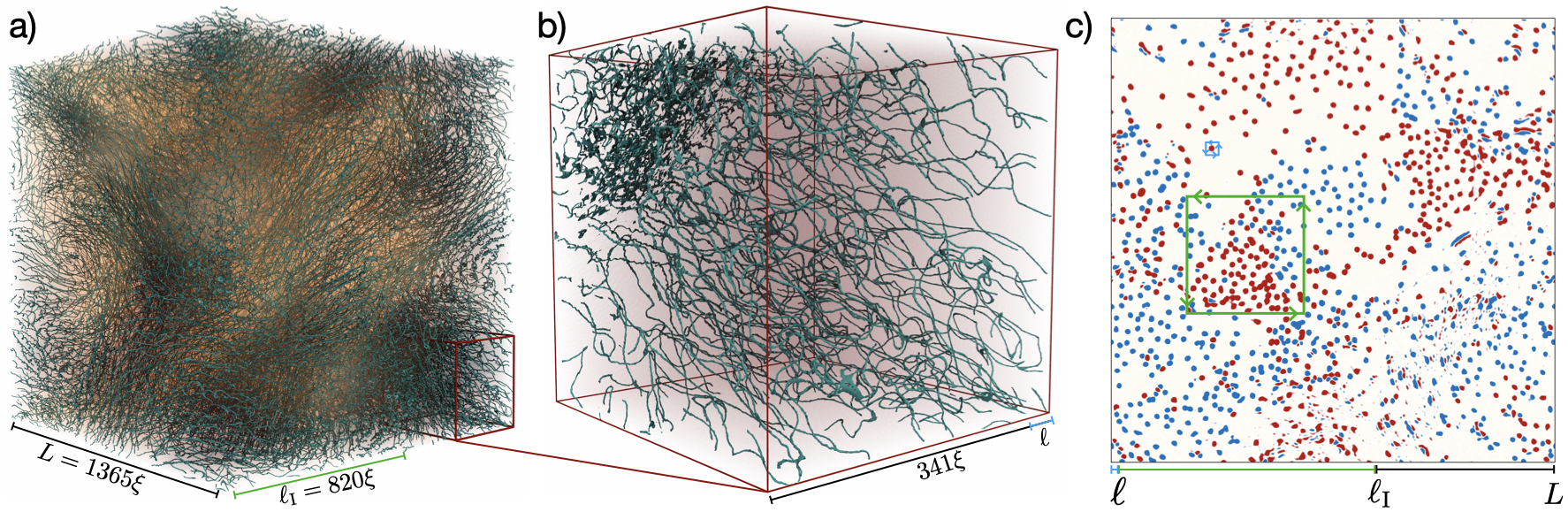}
  \caption[]{%
    Visualization of a quantum turbulent vortex tangle from gGP simulations using
    $2048^3$ collocation points.
    (a) Full simulation box.
    Quantum vortices are displayed as thin teal-colored filaments, and correspond to isosurfaces of a vanishingly small density value.
    Density fluctuations about its bulk value are volume-rendered in shades of brown.
    The size of the box $L$ is expressed in units of the healing length $\xi$, that is of the order of the vortex core size. $\Lint$ is the integral scale of the flow.
    (b) Zoom of the full box. The mean inter-vortex distance $\ell$ is indicated at the bottom of the figure. 
    (c) Two-dimensional slice of the full box displaying the low-pass filtered vorticity field. Blue and red dots correspond to vortices of different signs.
    Also shown are two typical integration loops: a small blue path surrounding a single vortex, and a larger green path enclosing several vortices.}\label{fig:visualisation}
\end{figure*}

Quantum flows are capable of reaching a turbulent state not unlike high Reynolds numbers classical flows.
Loosely speaking, quantum turbulence is described as a complex \emph{tangle} of quantum vortices, as illustrated by the teal-colored filaments in the flow visualization in Fig.~\ref{fig:visualisation} (see details on the numerical simulations later).
Such turbulent tangle displays rich multi-scale physics.
At scales larger than the mean distance between vortices $\ell$, the quantum nature of vortices is less dominant and fluid structures, akin to those observed in classical fluids, are apparent [Fig.~\ref{fig:visualisation}(a)]. In contrast, at scales smaller than $\ell$, the dynamics of individual quantized filaments becomes very important.
Figure~\ref{fig:visualisation}(b) displays a zoom of the flow, where Kelvin waves (waves propagating along vortices) and vortex reconnections are clearly observed.
Due to this multi-scale physics, with discrete vortices at small scales and a classical-like behavior at large ones, quantum turbulence can be considered as the skeleton of classical three-dimensional turbulent flows~\cite{Barenghi2014,Tsatsos2016}. Such ideas will be further supported by the results discussed later in this work.

Classical turbulent flows are characterized by an inertial range of scales where, according to the celebrated Kolmogorov's K41 theory~\cite{Kolmogorov1941a}, statistics are self-similar and independent of
the energy injection and dissipation mechanisms.
In particular, the variance of the velocity circulation is expected to follow the power law scaling $\mean{\CircA^2} \sim A^{4/3}$ when the loop area $A$ is within the inertial range.
This prediction, based on dimensional grounds, is equivalent to the two-thirds law for the variance of the Eulerian velocity increments~\cite{Frisch1995}.
The four-thirds scaling law for the circulation variance has been robustly observed in classical turbulence experiments~\cite{Sreenivasan1995a,Zhou2008} and numerical simulations~\cite{Umeki1993,Cao1996,Benzi1997,Iyer2019a}.
Furthermore, as shown by these studies, higher order circulation moments robustly deviate from K41 scalings.
Such deviations result from the intermittency of turbulent flows~\cite{Frisch1995,Sreenivasan1997}, that is, the emergence of rare events of extreme intensity, associated to the breakdown of spatial and temporal self-similarity.
Very recently, high Reynolds number simulations have shown that the intermittency of circulation may be described by a very simple bifractal model~\cite{Iyer2019a}, which contrasts with the more complex multifractal description of velocity increment statistics.
This study has renewed interest on the dynamics of circulation in classical turbulence~\cite{Apolinario2020,Migdal2020b,Iyer2020c}.

As in classical flows, K41 statistics and deviations due to intermittency have
been indeed observed in the large scales of quantum turbulence.
In particular, superfluid helium experiments have shown that finite-temperature
quantum turbulence is intermittent, and that the scaling exponents of
velocity increments might slightly differ from those in classical
turbulence~\cite{Maurer1998,Salort2011a,Rusaouen2017,Varga2018a}.
In zero-temperature superfluids, numerical simulations of the Gross--Pitaevskii
(GP) model have shown evidence of a K41 range in the kinetic energy
spectrum~\cite{Nore1997,Sasa2011,Muller2020a}.
Noting that the GP velocity field is compressible and singular at the vortex
positions, the energy spectrum is often computed using the incompressible part of a
regularized velocity field~\cite{Nore1997a}. This decomposition was used in Ref.~\cite{Krstulovic2016} to show that, in quantum turbulence, the intermittency of
velocity increments is enhanced with respect to classical turbulence.
Note that such decomposition is not needed for circulation statistics, since
the compressible components of the velocity are by definition potential
flows~\cite{Nore1997}, and therefore their contributions to the circulation
vanish when evaluating the contour integral in Eq.~\eqref{eq:circulation_intro}.
This absence of ambiguity, as well as its discrete nature, make the circulation a particularly interesting
quantity to study in low-temperature quantum turbulence.

The manuscript is organized as follows. In Section~\ref{sec:gGP_model} we present the model used in this work to simulate quantum turbulence, and we discuss the numerical methods to integrate it and to process data. Section~\ref{sec:circulation} presents and discusses the main results concerning the scaling of circulation moments in quantum turbulence and its intermittency. Finally, Section~\ref{sec:summary} discusses the implications of this work.

\section{Quantum turbulence simulations}%
\label{sec:gGP_model}

We numerically study the scaling properties of velocity circulation in quantum
turbulence. The results are obtained from a database of high-resolution direct numerical
simulations of a generalized Gross--Pitaevskii (gGP) model, which describes more
closely the phenomenology of superfluid helium compared to the standard GP
equation~\cite{Muller2020a}. The simulation reported in this work uses $2048^3$ grid points. In the following, we briefly introduce the gGP model used in this work.
For details, the reader is referred to Ref.~\cite{Muller2020a}.

The gGP equation writes
\begin{equation}
\begin{split}
  \label{eq:gGP}
  i \hbar \pdv{\psi}{t} &=
  - \frac{\hbar}{2m} \laplacian \psi
  - \mu (1 + \chi) \psi
  \\
  & + g \left(
    \int V_\text{I}(\bm{x} - \bm{y}) \, |\psi(\bm{y})|^2 \, \mathrm{d}^3 y
  \right) \psi
  + g \chi \frac{|\psi|^{2 (1 + \gamma)}}{n_0^\gamma} \psi,
\end{split}
\end{equation}
where $\psi$ is the condensate wave function describing the dynamics of a
compressible superfluid at zero temperature. Here, $m$ is the mass of the bosons, $\mu$ is the chemical potential, $n_0$ the particles density and
$g=4\pi\hbar^2a_s/m$ is the coupling constant proportional to the $s$-wave
scattering length.
To model the presence of the roton minimum in superfluid $^4$He, the governing equation includes a non-local interaction potential $V_\text{I}$ that is described in Appendix~\ref{app:nonlocal}.
This model also includes a beyond mean field correction controlled by two dimensionless parameters $\chi$ and $\gamma$ that correspond to its amplitude and order, respectively. This term arises from  considering a strong interaction between bosons~\cite{Muller2020a}.
Note that the standard Gross--Pitaevskii equation is recovered by setting $\chi = 0$ and $V_\text{I}(\bm{x}-\bm{y}) = \delta(\bm{x}-\bm{y})$, where $\delta$ is Dirac's delta.

The connection between Eq.~\eqref{eq:gGP} and hydrodynamics is given by the Madelung transformation, $\psi = \sqrt{\rho / m} \, e^{im\phi/\hbar}$, which relates $\psi$ to the velocity field $\vvec = \gradient \phi$. Note that the phase $\phi$ is not defined at the locations where the density $\rho$ vanishes, and hence the velocity is singular along superfluid vortices~\cite{Nore1997a}.
When the system is perturbed around a flat state $\psi=\sqrt{n_0}$, the speed of sound is given by $c = \sqrt{gn_0(1+\chi(\gamma+1))/m}$~\cite{Muller2020a}. Non-dispersive effects are observed at scales
below the healing length $\xi = \hbar / \sqrt{2mgn_0(1+\chi(\gamma+1))}$. This length scale is also the typical size of the
vortex core.

Equation~\eqref{eq:gGP} is solved using the Fourier
pseudo-spectral code FROST in a periodic cube with a fourth-order Runge--Kutta method for the time integration. In this work, the simulation box has a size $L=1365\xi$ and the initial condition is generated to follow the Arnold--Bertrami--Childress (ABC) flow used in Ref.~\cite{ClarkdiLeoni2017}. The initial velocity wave function is generated as a combination of two ABC flows at the two largest wave numbers as described in~\cite{Muller2020a}.
To reduce the acoustic emission, the initial condition is prepared using a minimization process~\cite{Nore1997a}.
Besides the integral length scale $\Lint$, associated with the largest scales of the initial condition, and the healing length, proportional to the vortex core size, in quantum flows it is possible to define a third length scale $\ell$ associated with the mean inter-vortex distance. This scale can be estimated as $\ell = \sqrt{L^3/\mathcal{L}}$ where $\mathcal{L}$ is the total vortex length of the system. Numerically, $\mathcal{L}$ is estimated using the incompressible momentum density as in Refs.~\cite{Nore1997a,Muller2020a}.

Evolving the initial setting under the gGP model [Eq.~\eqref{eq:gGP}] leads to the tangle of quantum vortices displayed in Fig.~\ref{fig:visualisation}, whose energetic content decays at large times as
vortices reconnect and sound is emitted~\cite{Villois2020}. Similar to decaying classical turbulence, this temporal decay is characterized by an intermediate stage, termed the \emph{turbulent} regime, in which the rate of dissipation of incompressible kinetic energy is maximal and the mean inter-vortex distance $\ell$ is minimal~\cite{Muller2020a}. In the present work we only consider this regime, as its large-scale dynamics is most comparable with fully developed classical turbulence. In this stage, as discussed in~\cite{Muller2020a}, the incompressible kinetic energy spectrum of high-resolution gGP simulations presents a clear K41 scaling range, followed by a Kelvin wave cascade range at small scales. At this time, the integral scale is measured to be $\Lint\approx820\xi$ and the inter-vortex distance $\ell\approx28\xi$ as illustrated in Fig.~\ref{fig:visualisation}.

Throughout this work, the circulation is computed from its velocity-based definition in Eq.~\eqref{eq:circulation_intro}, as opposed to the vorticity-based expression resulting from application of Stokes' theorem (see Appendix~\ref{app:circulation_computation}). Moreover, only planar square loops of area $A = r^2$ are considered. We will thus refer to the circulation over a loop of area $A$ as $\CircA$ or $\CircR$, depending on the context.
To take advantage of the spectral accuracy of the solver, the circulation is computed from the Fourier coefficients of the velocity field, as detailed in Appendix~\ref{app:circulation_computation}.
Moreover, to reduce spurious contributions from loops passing close to vortices, each two-dimensional slice where circulation is computed is resampled into a finer grid of resolution $32 768^2$, using Fourier interpolation. Values of circulation are then filtered to keep only multiples of $\kappa$. Details on this procedure are given in
Appendix~\ref{app:circulation_computation}.

\section{Scaling of circulation in quantum turbulence}%
\label{sec:circulation}

The quantization of circulation is one of the defining properties of
superfluids.
However, despite its relevance, the behavior of circulation at scales much
larger than the vortex core size $\xi$ (about an Ångström in superfluid $^4$He)
is currently poorly understood in quantum turbulence.
Figure~\ref{fig:visualisation}(c) displays a two-dimensional cut of the fluid where a low-pass filtered vorticity field is displayed.
Vortices are visible as small dots, and their sign is colored in black and red.
Intuitively, one can expect that the circulation will be allowed to take increasingly higher values as the area of the integration loop increases.
For sufficiently small loops (such as the small path displayed in Fig.~\ref{fig:visualisation}(c)) the probability of enclosing a quantum vortex (let alone many of them) is small, and the circulation will most likely take values in $\{0, \pm \kappa\}$.
This strongly discrete distribution of circulation is in stark contrast with the continuous distribution found in viscous flows.
For larger loops, of size typically larger than the mean inter-vortex distance $\ell$, higher circulation values become possible as more vortices may intersect the loop area, as displayed by the large green path in Fig.~\ref{fig:visualisation}(c).
Even though it remains quantized, the discreteness of circulation becomes less apparent as the set of possible values increases.
Other effects, such as the cancellation of circulation contributions from anti-polarized vortices, become important.
Indeed, the relative orientation of quantum vortices is deeply linked to the emergence of K41 statistics in quantum turbulence~\cite{Lvov2007,Baggaley2012a}, and is expected to play a major role on circulation statistics at large scales.
The polarization of vortices is manifest in Fig.~\ref{fig:visualisation}(c), where, at large scales, vortices of the same sign have tendency to cluster.

\subsection{Circulation at classical and quantum scales \label{sec:circulationVariance}}

\begin{figure}[b]
  \centering
  \includegraphics[width=\linewidth]{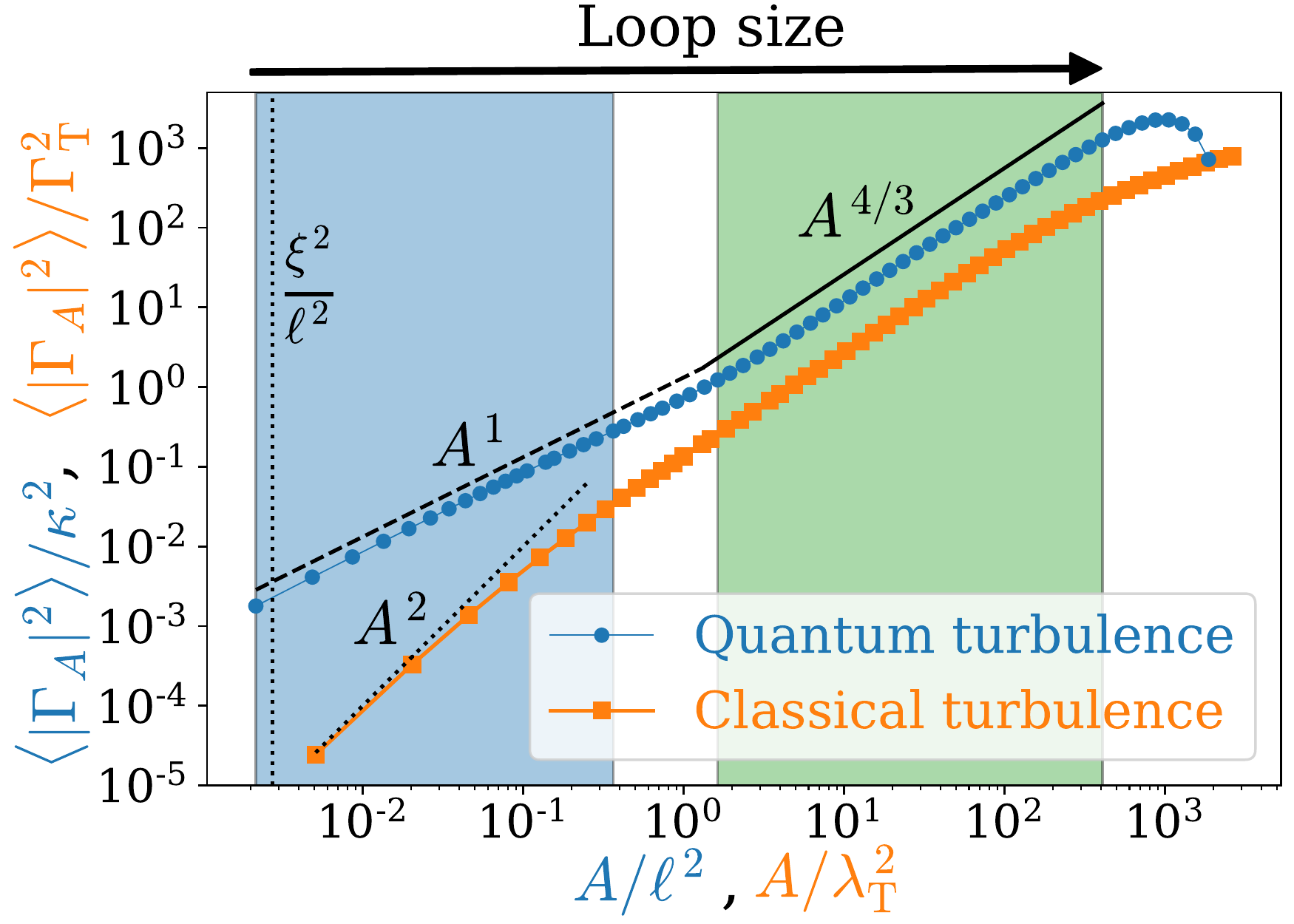}
  \caption{%
    Variance of the circulation about square loops of area $A = r^2$.
    Blue line shows the gGP simulation (resolution $ 2048^3$), and
    orange line the Navier--Stokes simulation (resolution $1024^3$).
    The classical variance is rescaled by $\Gamma_{\!\text{T}}^2 = \frac{\TaylorScale^4}{3} \mean{|\vortvec|^2}$, with $\TaylorScale$ the Taylor microscale and $\vortvec$ the vorticity field.
    }
    \label{fig:circulation_variance}
\end{figure}

We start by presenting one of the simplest circulation observables, that is the variance of the circulation for loops of different sizes in quantum turbulence. The scaling of the circulation with the area of the loops is displayed in Fig.~\ref{fig:circulation_variance}.
For comparison purposes, we also performed direct numerical simulations of the Navier--Stokes equations (see Appendix~\ref{app:navier_stokes}).
We then computed the scaling of the circulation variance in the steady state at a Taylor scale Reynolds number of $\ReLambda \approx 320$.
In the quantum flow, the circulation variance shows clear evidence of two scaling regimes. First, just like in the inertial range of classical turbulence, quantum turbulence displays a \emph{classical} range, where the $\mean{\CircA^2} \sim A^{4/3}$ scaling predicted by K41 theory is observed. This range corresponds to integration loops of linear dimension $r$ such that $\ell \ll r \ll \Lint$, where $\Lint$ is the integral scale of the flow.

In quantum turbulence, the emergence of K41 statistics for $r \gg \ell$ requires the partial polarization
of vortex filaments~\cite{Lvov2007,Baggaley2012a}, which effectively form bundles of
corotating vortices~\cite{Barenghi2014}.
For instance, due to vortex cancellations, a tangle of randomly
oriented vortices would be associated to $\mean{|\CircA|^2} \sim A$
in the classical range~\cite{Lvov2007}, different from the K41 estimate
$\mean{|\CircA|^2} \sim A^{4/3}$ verified in Fig.~\ref{fig:circulation_variance}.
On the other side of the spectrum, a fully polarized tangle (as may be found in
quantum flows under rotation) is associated to the estimate $\mean{|\CircA|^2} \sim A^2$.
We therefore see that K41 dynamics corresponds to a precise intermediate state between an isotropic and a fully polarized tangle.

At small scales, classical and quantum flows display different power law scalings.
Viscous flows are smooth at very small scales, and the vorticity field may be considered as constant within a sufficiently small loop.
By isotropy, it follows that $\mean{\CircA^2}\approx \mean{|\omega_iA|^2}=\frac{1}{3} \mean{|\vortvec|^2} A^2$ for small $A$.
Equivalently, such scaling can be obtained by invoking the smoothness of the velocity field and performing a Taylor expansion around the center of the loop~\cite{Iyer2019a}.
This viscous scaling is indeed observed in Fig.~\ref{fig:circulation_variance} for $r \ll \TaylorScale$.
Here, $\TaylorScale = v_{\text{rms}} / \sqrt{\mean{(\partial_x v_x)^2}}$ is the Taylor microscale, below which the dynamics of the flow is affected by viscosity in classical turbulence (see footnote~\footnote{%
  The Taylor micro-scale $\TaylorScale$ is formally defined from the longitudinal correlation function of the velocity field, as the scale at which its parabolic approximation at the origin vanishes~\cite{Pope2000}.
  It can be seen as the scale at which velocity gradients becomes important and viscosity starts to act.
  It is related to the Kolmogorov length scale $\eta$, the scale at which the turbulent cascade ends, by the relationship $\TaylorScale = 15^{1/4} \ReLambda^{1/2} \eta$, with $\ReLambda = v_{\text{rms}} \TaylorScale / \nu$ the Taylor scale Reynolds number~\cite{Frisch1995}.
  It is often used by experimentalists and theoreticians as it depends only on intrinsic properties of the turbulent flow, and not on the forcing and dissipative mechanisms}).
Note that we have used the Taylor microscale instead of the Kolmogorov length scale, which for the present numerical simulations is about thirty times smaller. This fact suggests that, in the correspondence between classical and quantum turbulence, the inter-vortex distance $\ell$ may be compared to the Taylor microscale.

On the contrary, for quantum turbulence, a less steep scaling is observed at small scales, which recalls the singular signature of the quantum vortex filaments. We will come back to this scaling later. In the following we refer to the range $\xi\ll r \ll \ell$ as the \emph{quantum} range, since it strongly differs from the dissipative range of classical turbulence. The quantum and  the classical ranges are highlighted by different background colors in Fig.~\ref{fig:circulation_variance}.
We have checked that the above results are also observed in low resolution simulations of the standard Gross--Pitaevskii model (data not shown).

\subsection{Circulation statistics and intermittency \label{sub:circulation_moments_and_intermittency}}%

\begin{figure}[t]
  \centering
  \includegraphics[width=\linewidth]{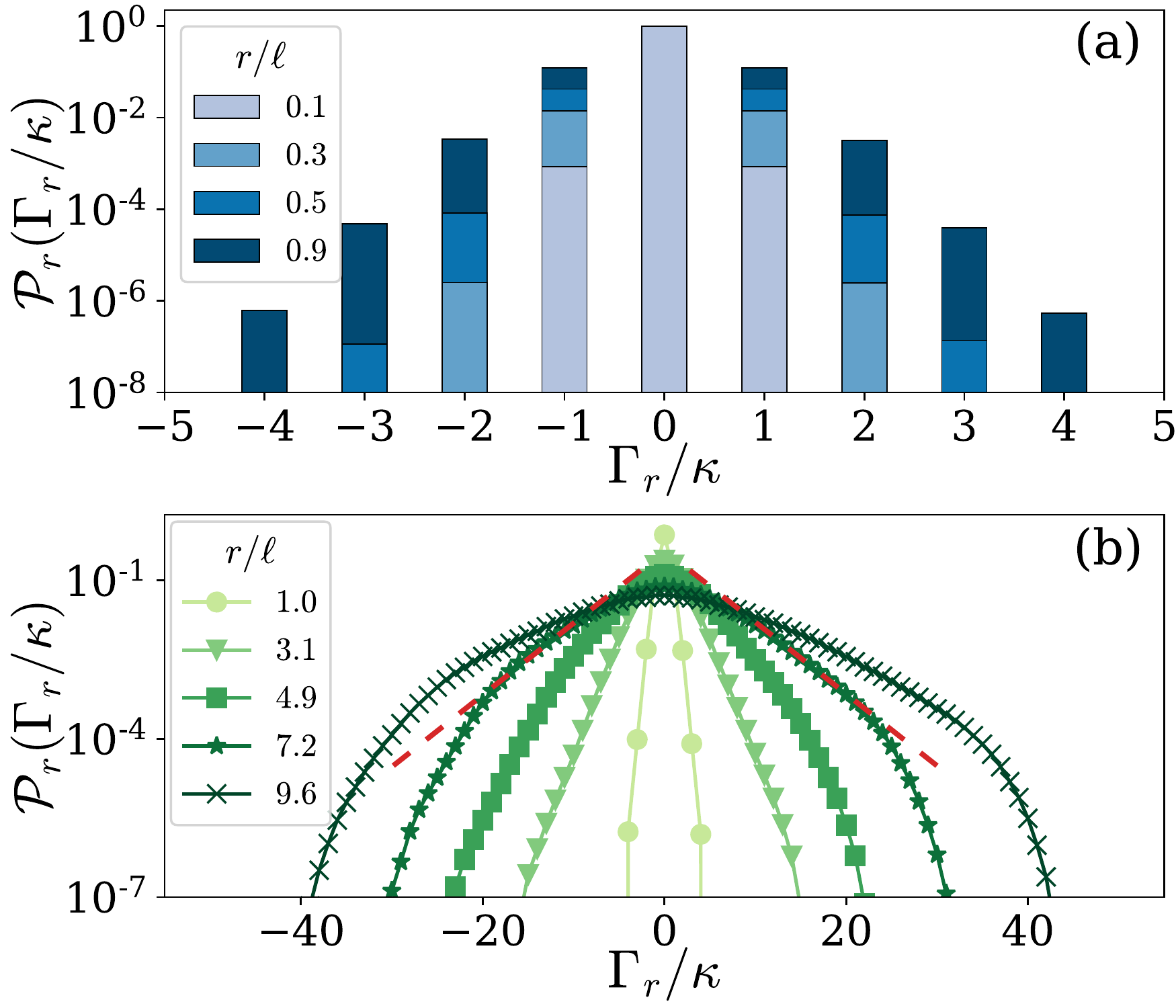}
  \caption{%
    Probability mass function (PMF) of the circulation in quantum turbulence for (a) loop sizes
  $r/\ell < 1$ and (b) loop sizes $r/\ell\geq1$.
  Red dashed lines are a guide for the eye indicating exponential tails.
  Note that all distributions are discrete, as $\CircR / \kappa$ only takes integer values.
  In (a), bars for $\Pcirc(0)$ and $r / \ell \ge 0.3$ are hidden behind the $r / \ell = 0.1$ case.
  See Fig.~\ref{fig:proba_zero_circ} for details on $\Pcirc(0)$.
}\label{fig:circulation_pdfs}
\end{figure}

In quantum flows, the velocity circulation takes discrete values (multiples of the quantum of circulation $\kappa$), which contrasts with the continuous space of possible values in viscous flows.
In statistical terms, its probability distribution is described by a probability mass function (PMF), the discrete analogous of a probability density function (PDF).
The discreteness of the circulation is most noticeable for loop sizes $r$ smaller than the mean inter-vortex distance $\ell$, where the probability of a loop
enclosing more than one vortex is vanishingly small, and $\CircR$ takes one of a small set of discrete values.
This is verified in Fig.~\ref{fig:circulation_pdfs}(a), where the probability $\Pcirc(n)$ of having a circulation $\CircR=n\kappa$, for small loop sizes, is shown.
As expected, the PMFs are strongly peaked at $\CircR = 0$ for very small loop sizes, indicating that it is very unlikely for such a loop to enclose more than one vortex (vortex cancellation is negligible at those scales).
The PMF becomes wider as $r$ increases and more vortices are allowed within an integration loop.

The circulation PMF within the quantum range strongly differs from the (continuous) PDF of circulation in the small scales of classical turbulence.
In isotropic flows, for a fixed loop size $r$ in the dissipative
range, the circulation PDF is equivalent to that of a vorticity
component.
Vorticity is a highly intermittent quantity in fully developed turbulence,
and as other small scale quantities, it is characterized by a strongly
non-Gaussian distribution with long tails~\cite{Wilczek2009}. In that sense, and in what concerns circulation, quantum turbulence presents a much simpler behavior despite its singular distribution of vorticity. Such a behavior could be useful for developing theoretical models of circulation.

For larger loops with $r / \ell > 1$, the circulation takes increasingly larger
values, and its discrete nature becomes less apparent.
This is seen in the circulation PMFs shown in
Fig.~\ref{fig:circulation_pdfs}(b), which may be approximated by continuous
distributions.
Within the classical range, these distributions seem to display
exponential-like tails (red dashed lines).
These distribution tails are compatible with those found in the inertial range of classical turbulence, which may be fitted by stretched exponentials~\cite{Iyer2019a} or modified exponentials~\cite{Iyer2020c}.

In classical turbulence, it is customary to characterize velocity intermittency
by evaluating the departure of the moments of velocity increments from K41
self-similarity theory~\cite{Frisch1995}.
For the same purposes, a few studies have also considered the moments of
circulation~\cite{Sreenivasan1995a,Cao1996,Benzi1997,Zhou2008,Iyer2019a,Iyer2020c}.
In the following, we consider the moments $\mean{|\CircR|^p}$ in quantum
turbulence resulting from the circulation distributions discussed in the
previous section.
The aims are to characterize the validity of K41 theory in the classical range,
provide evidence of possible departures due to intermittency, and to elucidate
the statistics of circulation at small scales resulting from the quantum nature
of the flow.
This extends the discussion relative to the circulation variance ($p = 2$),
presented in Fig.~\ref{fig:circulation_variance} in the context of a comparison
with classical flows.

\begin{figure}[b]
  \centering
  \includegraphics[width=\linewidth,height=1.1\linewidth]{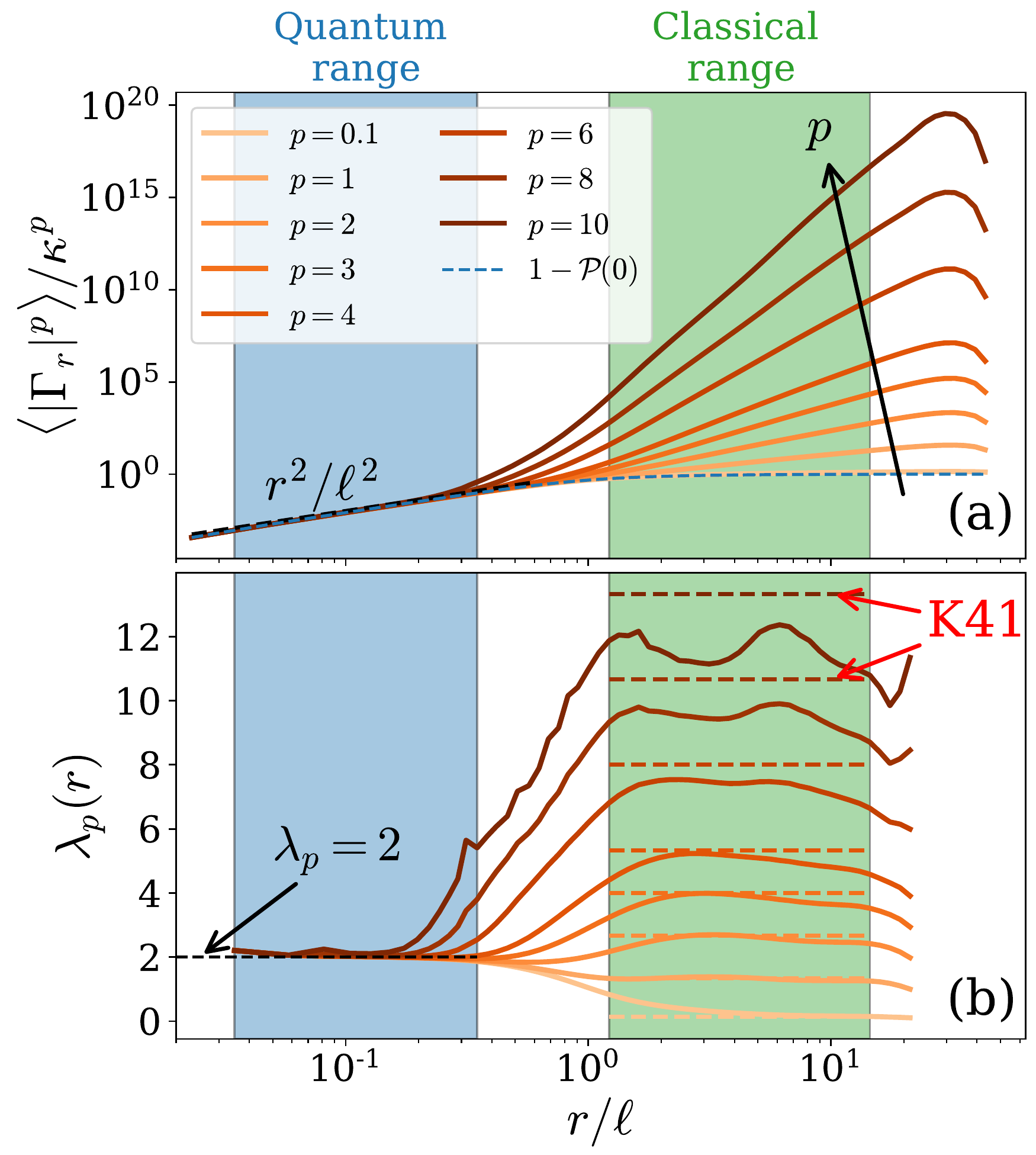}
  \caption[]{%
          (a) $p$-order moments of the circulation over square loops of area
          $A = r^2$ from the gGP simulation.
          (b) Local scaling exponents $\lambda_p(r) = \dd [\log
          \mean{|\CircR|^p}] / \dd [\log r]$.  Dashed horizontal lines
          correspond to the K41 scalings $\lambdaKolm_p = 4p/3$. The blue dashed
          line shows $\Pcirc(\CircR\neq0) = 1 - \Pcirc(\CircR=0)$,
          which corresponds to the $p$-th circulation moment in the limit $p
          \rightarrow 0^+$ [Eq.~\eqref{eq:Gamapzero}].  The blue and
          green areas respectively illustrate the quantum and classical
          regimes.
  }\label{fig:circulation_moments}
\end{figure}

Circulation moments $\mean{|\CircR|^p}$ are shown in
Fig.~\ref{fig:circulation_moments}(a) as a function of the loop size $r$ for different orders $p$.
For each moment, a clear power law scaling is identified in each of these
ranges. We define the exponents of the power law as
\begin{equation}
  \label{eq:moments_scaling}
  \frac{\mean{|\CircR|^p}}{\kappa^p} \approx
  {\left( \frac{r}{\ell} \right)}^{\! \lambda_p}.
\end{equation}
To better characterize the exponents, one can compute the local scaling exponents $\lambda_p(r) = \dd [\log \mean{|\CircR|^p}] /
\dd [\log r]$, that for pure power laws are flat. The local scaling exponents are presented in Fig.~\ref{fig:circulation_moments}(b), where two different plateaux are observed in both ranges for each order $p$.

\paragraph{Quantum range}

At first glance, it is striking to note that all moments collapse in the quantum
range.
This suggests that circulation is extremely intermittent at these scales as consequence of the quantum nature
of the flow.
Indeed, as inferred from Fig.~\ref{fig:circulation_pdfs} and discussed in the
previous section, a random loop of characteristic length $r \ll \ell$ will
almost never enclose more than a single vortex filament.
By the definition of the inter-vortex distance $\ell$, at such small scales, the probability of finding a vortex within a loop is simply $\beta_r=r^2/\ell^2$. From there it follows that $\mean{|\CircR|^p}=(0\times\kappa)^p(1-\beta_r)+(1\times\kappa)^p\beta_r$, since only zero or one vortex might lay inside the loop. This simple model leads to the prediction
\begin{equation}
  \label{eq:moments_quantum}
  \frac{\mean{|\CircR|^p}}{\kappa^p} \approx
  {\left( \frac{r}{\ell} \right)}^{\! 2}
  \qfor r \ll \ell,
\end{equation}
which is precisely the law observed in Fig.~\ref{fig:circulation_moments} at
small scales.
Remarkably, the simulations results capture not only the predicted scaling
exponent $\lambda_p = 2$ (as verified in Fig.~\ref{fig:circulation_moments}(b)),
but also the prefactor $\ell^{-2}$.

The independence of the circulation scaling exponents $\lambda_p$ on the moment
order $p$ translate the extreme intermittency of circulation at quantum scales.
This result is a clear consequence of quantum physics, as it results from the
quantization of circulation and the discrete nature of vortex filaments.
As seen in Fig.~\ref{fig:circulation_variance}, it is in stark contrast with the
small-scale physics of viscous flows, characterized by smooth velocity fields,
which lead to very different circulation statistics scaling as $r^{2p}$.

\paragraph{Classical range}

For larger loops of size $\ell \ll r \ll \Lint$, circulation moments in
Fig.~\ref{fig:circulation_moments} follow different power laws, with a scaling
exponent $\lambda_p$ that increases with the moment order $p$.
Kolmogorov's phenomenology gives a prediction for the scaling of circulation moments in this regime. Assuming self-similarity across scales, the K41 predictions for the circulation moments about loops of area $A = r^2$ are of the form
\begin{equation}
  \label{eq:moments_K41}
  \mean{|\CircR|^p} = C_p \, \varepsilon^{p/3} r^{4p/3}
  \qfor \ell \ll r \ll \Lint
\end{equation}
for positive moment order $p$, where $\varepsilon$ is the incompressible kinetic
energy dissipation rate per unit mass, and $C_p$ are supposedly universal constants.
Similarly to classical K41 scalings, Eq.~\eqref{eq:moments_K41} results from dimensional arguments and the assumption that, within the classical range, the statistics of $\CircR$ depends only on $\varepsilon$ and $r$.

The local scaling exponents displayed in Fig.~\ref{fig:circulation_moments}(b) exhibit a plateau in the classical range,
confirming the power-law behavior of circulation moments at those scales.
For low order moments ($p<3$), the exponents approximately match the
K41 prediction, plotted as dashed horizontal lines.
This observation is consistent with the scaling of the circulation variance in
Fig.~\ref{fig:circulation_variance}.
On the other hand, higher order moments yield lower exponent values than those
predicted by K41 theory.
This departure is clear evidence of circulation intermittency in the classical
range of quantum turbulence.
Moreover, it is qualitatively consistent with the trends observed in the
inertial range of classical
turbulence~\cite{Sreenivasan1995a,Zhou2008,Cao1996,Benzi1997,Iyer2019a}.
A more quantitative comparison of the scaling exponents in classical and quantum
flows is provided in the next section.

\subsection{Scaling exponents in the classical regime}

We finally quantify the anomalous exponents of the circulation in the classical
range of the quantum turbulent tangle.
For this, we average the local scaling exponents over a range of loop sizes
within $\ell \ll r \ll \Lint$.
The precise averaging range is given by the green area in
Fig.~\ref{fig:circulation_moments}.
As in Ref.~\cite{Iyer2019a}, we also compute fractional circulation
moments.
However, note that we do not include negative moments $p \in \, ]-1, 0[$ as done
in that work, because due to the discrete nature of the circulation distribution
in quantum flows, there is a finite probability of having $\CircR=0$, and thus negative order moments diverge.

\begin{figure}[b]
  \centering
  \includegraphics[width=\linewidth]{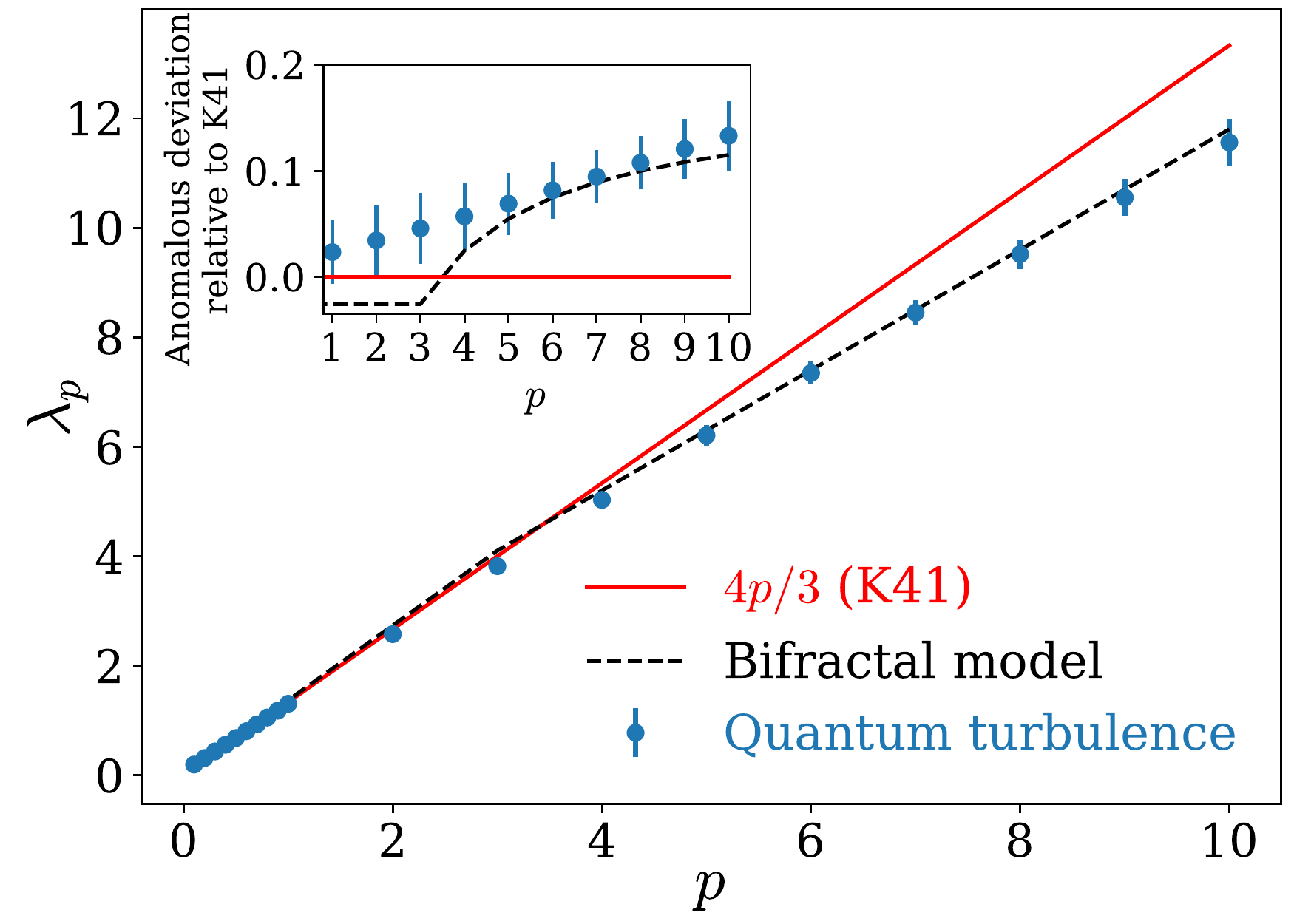}
  \caption[]{%
    Scaling exponents of the circulation moments for loop sizes within the
    classical range ($\ell \ll r \ll \Lint$).
    Blue circles with error bars correspond to gGP simulations;
    solid line, K41 scaling $\lambdaKolm_p = 4p/3$;
    dashed line, bifractal fit in classical turbulence~\cite{Iyer2019a}.
       Inset: relative deviation from K41 estimates, $(\lambdaKolm_p - \lambda_p) /
    \lambdaKolm_p$.
    Error bars indicate the standard deviation of each $\lambda_p$ within the classical
    range.
  }\label{fig:scaling_exponents}
\end{figure}

The circulation scaling exponents $\lambda_p$ obtained from our simulations are
shown in Fig.~\ref{fig:scaling_exponents}.
As suggested by the behavior of the circulation moments discussed in the previous section, the
departure from K41 scaling (solid red line in the figure) is weak for low order
moments, while it becomes significant for orders $p \ge 3$.

Strikingly, the scaling exponents are consistent with the recent results in high
Reynolds number classical turbulence~\cite{Iyer2019a} (dashed lines in
Fig.~\ref{fig:scaling_exponents}).
To give some relevant context, that work provides evidence of a bifractal
behavior of the scaling exponents.
Concretely, for low order moment $p < 3$, the exponents grow linearly as
$\lambda_p = \alpha p$ with $\alpha \approx 1.367$.
This robust scaling, almost independent of Reynolds number, is close but not
exactly equal to the $\alpha = 4/3$ predicted from K41 phenomenology.
As for orders $p > 3$, they are accurately described by a monofractal fit
$\lambda_p = hp + (3 - D)$, with a fractal dimension $D$ and Hölder exponent $h$
that display a weak Reynolds number dependence.
At the highest Reynolds number studied in that work, these are estimated as
$D \approx 2.2$ and $h \approx 1.1$.
We stress that the above bifractal fit, that we adopt here for its simplicity, is empirically derived in~\cite{Iyer2019a} from direct numerical simulation data.
Note that an alternative functional form of the scaling exponents $\lambda_p$ in classical turbulence, which also closely matches the numerical data, has been recently proposed based on a dilute vortex gas model~\cite{Apolinario2020}.

For high order moments, the anomalous exponents in the quantum flow case display a behavior that is close to that observed in classical turbulence. The inset of Fig.~\ref{fig:scaling_exponents} shows the relative deviation from K41 estimates, $(\lambdaKolm_p - \lambda_p) / \lambdaKolm_p$, and its comparison with the bifractal model fitted in~\cite{Iyer2019a}. For $p>3$, the bifractal model lays between error bars of our data. This hints towards the universality of inertial range dynamics across different
turbulent systems.

Low order moments are particularly interesting. From a statistical point of view, the main contribution to those moments comes from loops having a very small circulation, which are the most probable ones (see Fig.~\ref{fig:circulation_pdfs}). A loop with small circulation might either be the result of a region of the flow where there are few vortices, or the opposite regime, where many vortices of opposite signs cancel each other's contributions to the circulation. The last case corresponds to a very rare intermittent event. Such idea was invoked by \textcite{Iyer2019a} to explain the intermittency of low order moments.

\begin{figure}[tb]
  \centering
  \includegraphics[width=\linewidth]{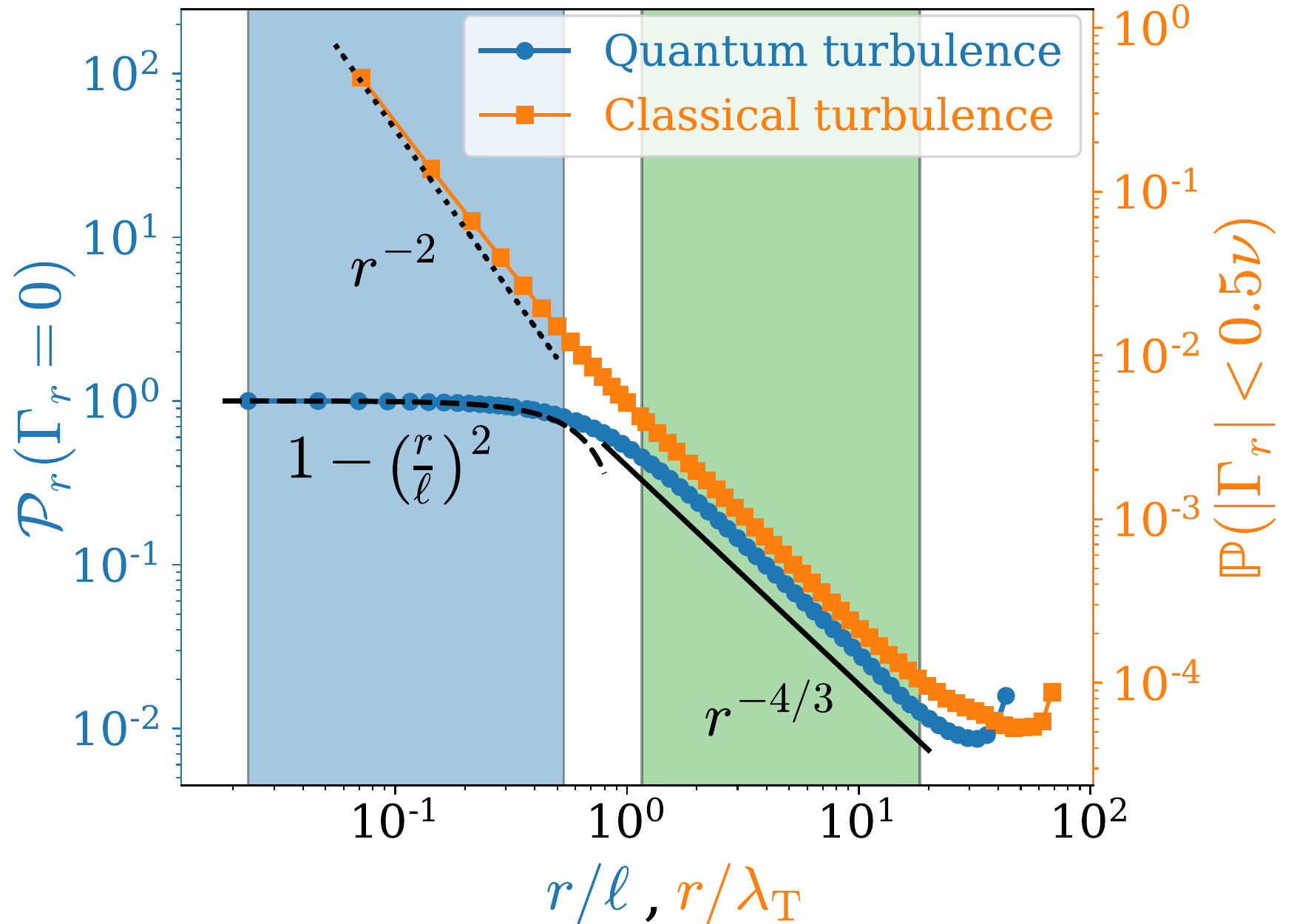}
  \caption[]{%
          Probability of having zero circulation in gGP simulations (blue line) and of having a weak circulation in Navier--Stokes simulations (orange line). Dashed lines show their respective predictions at large and small scales.
  }\label{fig:proba_zero_circ}
\end{figure}

In the case of quantum turbulence, the discrete nature of vortices is very important, and regardless of the size of the loop, there is always a non-zero probability of having a total zero circulation. In fact, we can relate low-order moments with such probability as
\begin{equation}
  \mean{|\CircR|^p} = \sum_{n\neq0} |\CircR|^p \, \Pcirc(n) = 1 - \Pcirc(0)+p\mean{\log{|\CircR|}}_{\neq0}+o(p),
  \label{eq:Gamapzero}
\end{equation}
where $\Pcirc(n)$ is the circulation PMF and $\mean{\mathcal{O}[\CircR]}_{\neq0} = \sum_{n\neq0} \mathcal{O}[\CircR] \, \Pcirc(n)$.
This results from the Taylor expansion $|\CircR|^p = 1 + p \log |\CircR| + o(p)$ about $p = 0$, and the fact that $\mean{1}_{\neq 0} = 1 - \Pcirc(0)$.
Remarkably, the probability of having zero circulation displays a clear $r^{-4/3}$ power law scaling in the classical regime, as shown in Fig.~\ref{fig:proba_zero_circ}.
This power law is related to a partial polarization of the quantum vortices. Indeed, in the case of a fully polarized tangle, we trivially have that $\Pcirc(0)=0$, as all vortices have the same sign within a loop. In the opposite regime of a totally unpolarized tangle, we have that $\Pcirc(0)\sim r^{-1}$. This scaling results from considering $N\sim (r/\ell)^2$ homogeneously distributed uncorrelated vortices enclosed in a loop of size $r$, and computing the probability of having exactly $N/2$ positive vortices among those $N$.
Such probability is simply given by
$2^{-N} \binom{N}{N/2} \approx \sqrt{2/N\pi}\sim (r/\ell)^{-1}$.
The $r^{-4/3}$-scaling thus corresponds to a partial polarization of the tangle. Note that the transition between the quantum and the classical regimes is manifest. At small scales, we find that $\Pcirc(0)=1-(r/\ell)^2$, which corresponds to the probability of not finding any vortex.

It is interesting that for classical flows, albeit the circulation takes continuous values, the probability $\Prob(|\CircR| < \alpha \nu)$ of having low circulation values presents the same power law in the inertial range, as also reported in Fig.~\ref{fig:proba_zero_circ}. For a classical flow, this scaling can be derived by invoking K41 phenomenology, which predicts that the statistics of $\gamma=\CircR\epsilon^{-1/3}r^{-4/3}$ is scale invariant in this range.
It follows that
\begin{equation}
  \Prob(|\CircR|<\alpha \nu) =
  \Prob(|\gamma|<\alpha \nu \epsilon^{-1/3}r^{-4/3})\sim \alpha \nu \epsilon^{-1/3}r^{-4/3}
\end{equation}
for $\alpha\ll1$. Here we assumed that the PDF of $\gamma$ is finite at zero. Besides, for $r$ much smaller than the Taylor microscale $\TaylorScale$, one has that $\CircR\sim \omega_i r^2$ (see Section~\ref{sec:circulationVariance}) and a similar argument leads to $\Prob(|\CircR|<\alpha \nu)\sim r^{-2}$, as is also displayed in Fig.~\ref{fig:proba_zero_circ}~\footnote{The previous discussion suggests to normalize distances using the Kolmogorov length $\eta=(\nu^3/\epsilon)^{1/4}$ instead of the Taylor microscale $\TaylorScale$.
However, for the sake of simplicity and consistency with Fig.~\ref{fig:circulation_variance}, we used $\TaylorScale$ in Fig.~\ref{fig:proba_zero_circ}.}. Again, the small scales of classical and quantum fluids strongly differ.

Finally, note that the asymptotic approach predicted in Eq.~\eqref{eq:Gamapzero} is clearly verified in Fig.~\ref{fig:circulation_moments} for low order moments. The finite value of $\Pcirc(0)$ in the quantum case implies a discontinuity of the moments when $p\to0^+$, since $\mean{|\CircR|^0}=1$. The subdominant power-law term in Eq.~\eqref{eq:Gamapzero} explains the reduced inertial range observed in Fig.~\ref{eq:moments_K41} for low order moments.

\section{Summary and discussion\label{sec:summary}}

The recent work of \textcite{Iyer2019a} has sparked renewed interest on the
statistics of velocity circulation in high Reynolds number classical turbulent
flows.
Their numerical results have showcased the relative simplicity of circulation
statistics in the inertial range, despite the intermittency of these flows.
This simplicity contrasts with the complexity of velocity increment statistics,
as well as those of enstrophy or dissipation, which display
multifractal statistics as a result of turbulence intermittency~\cite{Frisch1995}.

It has been long suggested that quantum turbulence shares many similarities with
classical flows at scales much larger than those associated to individual
quantum vortices.
For instance, experimentalists have struggled to find significant differences
between finite-temperature superfluid helium and classical
flows at those scales~\cite{Maurer1998,LaMantia2013a,Rusaouen2017}.
Features of classical turbulence, most notably the scaling of the energy
spectrum $E(k) \sim k^{-5/3}$ resulting from Kolmogorov's self-similarity
theory, have also been observed in low-temperature quantum
turbulence~\cite{Nore1997a,Araki2002,Parker2005,Kobayashi2005,Kobayashi2007,Yepez2009,Tsubota2009,Sasa2011,Barenghi2014a,ClarkdiLeoni2017}.
However, for a few reasons detailed below, such observations only show a limited picture of
inertial range dynamics in quantum flows.
Firstly, most of these studies have looked at the scaling properties of the
velocity field and its wave number spectrum.
The velocity field is a singular quantity which diverges at the vortex filament
locations.
This has lead to considering a regularized version of it, whose physical
interpretation is less clear.
Secondly, even though K41 scaling has been observed in low-temperature
quantum turbulence, little is known regarding deviations from them due to
intermittency.
Indeed, despite a few works~\cite{Krstulovic2016,Rusaouen2017,Varga2018a}, because of numerical of experimental limitations, non-conclusive results exist on how the intermittency of those flows compares with classical turbulence.
Due to having two disjoint ranges of scales with non-trivial dynamics (as
opposed to just one in classical turbulence), high resolutions are needed to
obtain more than a decade of inertial range in wave number space~\cite{Tsubota2017,ClarkdiLeoni2017,Muller2020a}.

The differences between classical and quantum turbulence become more evident at
smaller scales, as the regularity of classical flows at scales below the
dissipative length is in stark contrast with the singular nature of quantized
vortices.
At those scales, quantization leads to enhanced intermittency of velocity
statistics in superfluid helium~\cite{Varga2018a,Biferale2018} and in
zero-temperature quantum turbulence~\cite{Krstulovic2016}.
Note that at quantum scales, both the singularity of the velocity field and
compressible effects such as sound emission become important.
As mentioned above, this leads to the necessity of regularizing and decomposing
the velocity field into different contributions.
In contrast, the velocity circulation considered in this work does not suffer
from such limitations, as it is non-singular, and, by
its definition, is exempt of contributions from compressible dynamics.

In this work we have numerically investigated circulation statistics in low
temperature quantum turbulence.
In superfluid flows, the velocity circulation is intimately linked to the quantum nature of the
system.
We have performed high resolution numerical simulations of a generalized
Gross--Pitaevskii model, allowing for a relatively large degree of scale
separation between the vortex core size $\xi$, the mean inter-vortex distance
$\ell$, and the integral scale of the flow $\Lint$.
The main objectives of this work have been twofold: (1) to disentangle the
differences between classical and quantum turbulence at small scales, and (2) to
provide new evidence of the strong analogy between both physics at large scales,
which, as we show, goes beyond self-similarity predictions and includes
intermittent behavior.
Our results strongly reinforce the view of quantum turbulence as the skeleton of
classical flows, which can be used to provide a better understanding of the latter. Besides, note that the physics of the Kelvin wave cascade, which becomes important at quantum scales, should play no role on circulation statistics, as the circulation around a vortex is blind to the presence of such vortex excitations.

We have considered the circulation $\CircR$ integrated over square loops of
varying area $A = r^2$.
As customary in classical turbulence, we have characterized the scaling properties of the circulation in terms of its moments $\mean{|\CircR|^p}$, and their dependence on the scale $r$ of the integration loop.
We have shown that all circulation moments follow two distinctive power law scalings, for $r$ much smaller and much larger than the mean inter-vortex distance $\ell$.

At small (or \emph{quantum}) scales, our main finding is that circulation
moments are independent of the moment order $p$, which translates the extreme
intermittency of the circulation at these scales.
This result is a consequence of the quantized nature of circulation and the
discreteness of vortex filaments. The small-scale dynamics of circulation in quantum flows is in strong contrast
with that in classical flows, where as a result of viscosity, the velocity field
is smooth at very small scales, leading to very different circulation
statistics.

At scales larger than $\ell$ (the \emph{classical} range), we have found that
low-order circulation moments closely follow the predictions of K41
phenomenology theory, which were initially proposed by Kolmogorov for classical
turbulence.
This is by itself a very important result, as it highlights the strong analogy
between classical and quantum flows at large scales.
While K41 scalings have been previously observed in the energy spectrum of
zero-temperature quantum turbulence, this is the most convincing evidence to
date of such behavior, as the circulation is a well defined physical quantity in
quantum turbulence, and the observed K41 range spans about one full decade in
scale space.

In addition, our work provides unprecedented evidence of intermittency in the
classical range of zero-temperature quantum turbulence.
The circulation moments obtained from our simulations not only display
intermittent behavior (in the form of deviation from K41 estimates), but they do
so in a way that is quantitatively similar to the anomalous scaling of circulation
in classical turbulence.
The impressive similarity between these two a priori very different systems,
strongly reinforces the idea of universality of inertial range dynamics in
classical and quantum flows.
Indeed, since Kolmogorov's pioneering works in 1941, it has been conjectured
that such dynamics is independent of the viscous dissipation mechanisms in
classical fluids.
The present work goes further to suggest that, more generally, inertial range
dynamics and intermittency are independent of the small-scale physics, and in particular, on the regularization mechanism.
In classical turbulence, viscosity has the role of smoothing out (or regularizing) the flow at small scales.
In quantum flows, regularization results from dispersive effects taking place at scales smaller than the vortex core size. Note that in~\cite{Chen2006}, it was suggested that for classical flows in the limit of infinite Reynolds numbers, the Kelvin theorem is violated and might be recovered only in a statistical sense, somehow as a consequence of the dissipative anomaly of turbulence~\cite{Frisch1995}. It would be of great interest to study how this picture changes in quantum turbulence, and to investigate whether an analogous of the classical circulation cascade exists~\cite{Eyink2006}.

In previous classical turbulence experiments~\cite{Sreenivasan1995a,Zhou2008}, circulation has been evaluated using the particle image velocimetry (PIV) technique, which provides a measure of the velocity field over a two-dimensional slice of the flow.
While this technique has been applied in finite temperature superfluid $^4$He~\cite{Donnelly2002,Zhang2005,Sergeev2009}, the interpretation of PIV measurements in this system remains unclear~\cite{Chagovets2011,Guo2014}.
As an alternative, variants of the particle tracking velocimetry (PTV) technique have been used in most recent studies of $^4$He~\cite{Chagovets2011,LaMantia2012,Guo2014,LaMantia2014a,Meichle2014,Marakov2015,Guo2019,Kubo2019,Wen2020}.
To our knowledge, no attempts have been made to compute the velocity circulation in superfluid experiments.
While perhaps challenging, such a study would be of large interest to the turbulence community.

The emergence of K41 scalings in quantum turbulence results from the partial polarization of vortex filaments~\cite{Lvov2007,Baggaley2012a}.
In quantum flows, due to the discrete nature of circulation, there is always a finite probability of having zero circulation, whose scale dependence also results from partial polarization.
Such a behavior is also seen in classical flows and can be explained by invoking K41 phenomenology.
This observation suggests that a possible stochastic modeling of classical and quantum turbulence, or at least of circulation statistics, could be based on a discrete combinatorial approach where spin-like vortices are generated with ad-hoc correlations. For such study, it will be important to gain a better understanding of the polarization of quantum turbulent tangles and of how this translates to classical flows.
Alternatively, in \textcite{Iyer2019a}, the bifractal behavior of circulation intermittency has been related to the presence of ``moderately wrinkled vortex sheets'' having a fractal dimension $D = 2.2$. It would be interesting to relate these ideas to the partial polarization and the arrangement of quantum vortices. Such ideas will be addressed in a future work.

\begin{acknowledgments}
  The authors thank L.\ Galantucci and U.\ Giuriato for fruitful discussions.
  This work was supported by the Agence Nationale de la Recherche through the project GIANTE ANR-18-CE30-0020-01. GK was also supported by the Simons Foundation Collaboration grant ``Wave Turbulence'' (Award ID 651471).
  This work was granted access to the HPC resources of CINES, IDRIS and TGCC under the allocation 2019-A0072A11003 made by GENCI.
  Computations were also carried out at the Mésocentre SIGAMM hosted at the Observatoire de la Côte d'Azur.
\end{acknowledgments}

\appendix

\section{Computation of circulation}%
\label{app:circulation_computation}

Via Stokes' theorem, Eq.~\eqref{eq:circulation_intro} about a closed loop $\Cloop$ can be written in terms of the vorticity field $\vortvec = \gradient \times \vvec$,
\begin{equation}
  \label{eq:app:circulation_vorticity}
  \CircA
  = \iint_A \vortvec \cdot \vb{n} \, \dd S,
\end{equation}
where $A$ is the area enclosed by the loop and $\vb{n}$ its associated normal unit vector.
Since the superfluid is irrotational away from vortices, this alternative form amounts to counting the contributions of the vortices enclosed within a loop.
In quantum flows, the vorticity field is extremely irregular, being effectively represented by a sum of Dirac deltas.
This renders Eq.~\eqref{eq:app:circulation_vorticity} impractical for numerically evaluating the circulation in quantum flows.

For the above reasons, we compute the circulation in quantum and classical flows using its velocity-based form Eq.~\eqref{eq:circulation_intro}.
The algorithm, described in the following, enables the evaluation of the line integral in Eq.~\eqref{eq:circulation_intro} with high accuracy over rectangular loops aligned with the Cartesian axes of the domain.
For simplicity, we consider a square loop of size $r \times r$, with sides respectively aligned with the $x$ and $y$ coordinate axes in a $2\pi$-periodic domain.
Here we denote by $\vvec(\xvec) = (v_x(x, y), v_y(x, y))$ the in-plane velocity field.

The circulation over such a square loop with opposite corners at $(x_0, y_0)$ and $(x_1, y_1) = (x_0 + r, y_0 + r)$ is given by
\begin{equation}
  \CircR =
    {\left[ V_x(y_0) \right]}_{x_0}^{x_1}
  + {\left[ V_y(x_1) \right]}_{y_0}^{y_1}
  - {\left[ V_x(y_1) \right]}_{x_0}^{x_1}
  - {\left[ V_y(x_0) \right]}_{y_0}^{y_1},
\end{equation}
where ${\left[ V_x(y) \right]}_a^b = \int_a^b v_x(x', y) \, \dd x'$ is the integral of $v_x$ along the $x$ direction.
This, and what follows below, similarly applies to the $y$ component of the velocity.

Using the Fourier representation of the velocity field, its $v_x$ component can be written as $v_x(x, y) = \sum_{k} \hat{u}_k(y) \, e^{i k x}$.
Then, its integral is ${\left[ V_x(y) \right]}_a^b = (b - a) \, \hat{u}_0(y) + \sum_{k \ne 0} \left[ -\frac{i}{k} \hat{u}_k(y) \right] \left( e^{i k b} - e^{i k a} \right)$.
However, note that the velocity field is singular at vortex locations, and as a result, the Fourier coefficients $\hat{u}_k$ decay slowly with the wave number $k$.
Hence, compared to the complex wave function $\psi$, a large number of Fourier modes is needed to accurately describe the velocity field.

In practice, to obtain an accurate representation of the velocity field on a given 2D cut of the 3D domain, we first evaluate the wave function $\psi(\xvec)$ on a 2D grid that is $\resampling$ times finer, along each direction, than the original $2048^2$ grid.
This evaluation is performed exactly from the Fourier coefficients of $\psi$.
In practice, this is done by zero-padding the Fourier representation of $\psi$ (from $2048$ to $2048 \resampling$ Fourier modes along each direction)\@.

\begin{figure}[b]
  \centering
  \includegraphics[width=\linewidth]{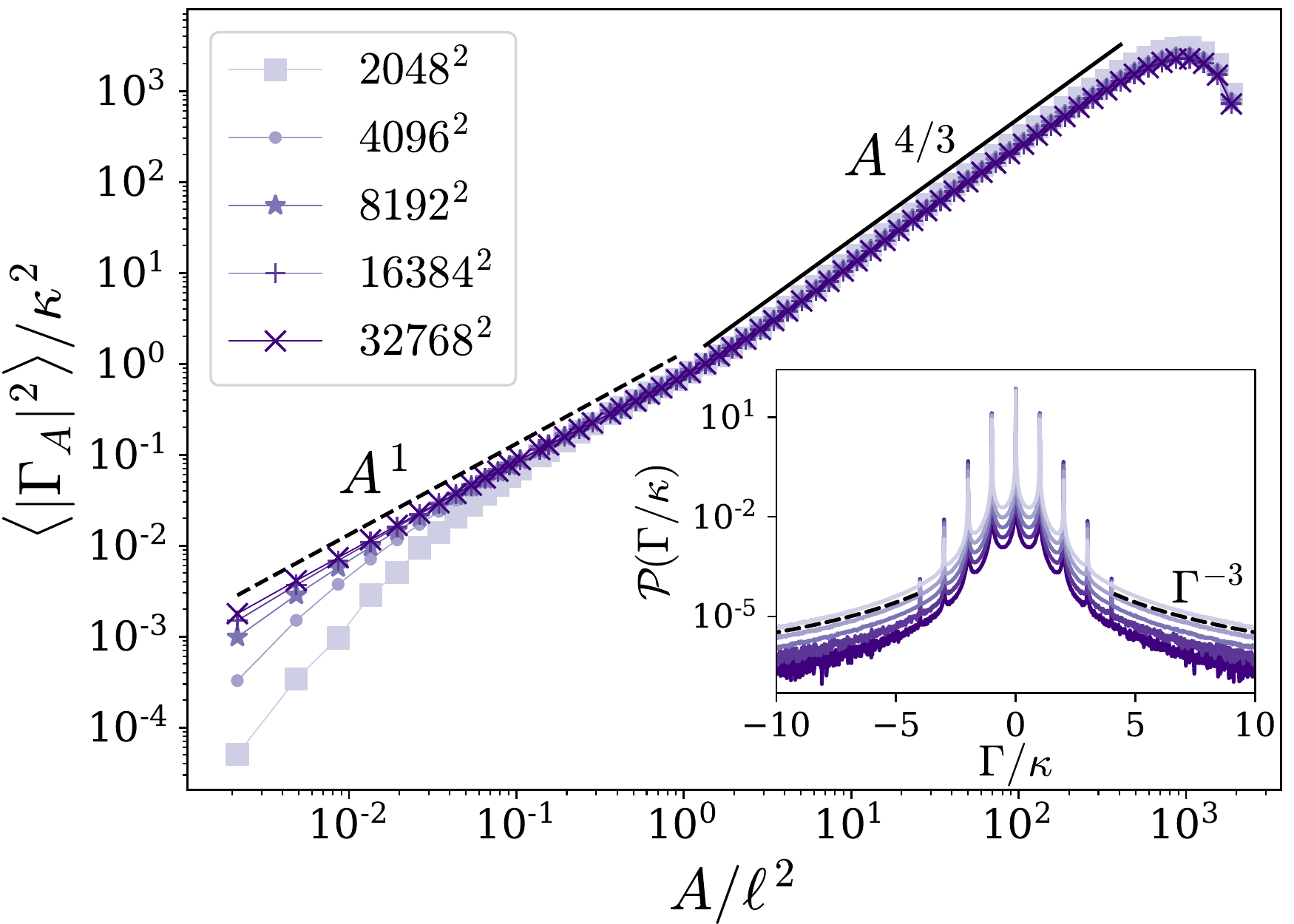}
  \caption{%
          Moments of order 2 for different values of the resampling in gGP simulations with $N^3=2048^3$.
          Resampling factors are $\resampling =$ 1, 2, 4, 8 and 16.
          As the resampling increases, vortices are better resolved and the expected scaling at small scales arises. At large scales the system is less affected by resampling. Inset: probability distribution of the circulation for a loop size $r/\ell = 0.67$. Peaks are observed at small circulation values multiples of $\kappa$. The separation between peaks and valleys is higher as the resampling increases. Tails exhibit a $\CircR^{-3}$ scaling.
  }\label{fig:gamma_variance_resampling}
\end{figure}

In Fig.~\ref{fig:gamma_variance_resampling}, we present the variance of the velocity circulation obtained using different values of the resampling factor $\resampling$.
For small loop sizes, the scaling $\mean{|\CircA|^2} \sim A^1$ predicted by Eq.~\eqref{eq:moments_quantum} is only observed when $\resampling$ is large enough ($\resampling \ge 8$), while for small $\resampling$, the small scale moments are contaminated by spurious circulation values.
Throughout this work, the value $\resampling = 16$ is used, i.e., the velocity is computed on a $\num{32768}^2$ grid for each 2D cut.
Note that for loop sizes in the classical range (where the K41 scaling $\mean{|\CircA|^2} \sim A^{4/3}$ is observed), resampling becomes less important.

Finally, the inset of Fig.~\ref{fig:gamma_variance_resampling} shows the measured PDF of the circulation along loops of size in the quantum range, for the same values of $\resampling$.
In all cases, the PDFs display peaks at small integer values of $\CircR / \kappa$, as expected from the underlying physics.
However, intermediate non-integer values are also sampled in the distributions.
These are a purely numerical artifact, mainly consequence of the approximation error arising from the Fourier truncation of the velocity field.
This error strongly decreases at high resampling factors, as evidenced by the increasing separation between peaks and valleys as $\beta$ increases.
Another source of spurious circulations originates when vortices are present very close to an integration path.
This leads to unphysical, very large circulation values sampling the $r^{-1}$ divergence of the velocity, whose signature are PDF tails exhibiting a $\CircR^{-3}$ scaling. As seen in the figure, resampling also helps reducing this error by a few orders of magnitude.
At a second step, these spurious contributions to the circulation distributions are further suppressed by only considering the peaks of $\CircR/\kappa$ close to integer values, from which discrete PMFs are constructed.
Only peaks which have a prominence of at least three orders of magnitude are considered, i.e., the value of the peaks should be at least 1000 times larger than its neighbors.

\section{Non-local interaction potential \label{app:nonlocal}}

To model the presence of the roton minimum in superfluid $^4$He, the governing
equation includes an isotropic non-local interaction potential~\cite{Reneuve2018,Muller2020a}
\begin{equation}
        \hat{\Vnl}(\bm{k}) = \left[ 1-V_1 \left(
        \frac{k}{k_{\mathrm{rot}}}\right)^2 + V_2 \left(
        \frac{k}{k_{\mathrm{rot}}}\right)^4 \right]
        \exp \left(-\frac{k^2}{2k_{\mathrm{rot}}^2} \right),
        \label{eq:nonlocal}
\end{equation}
where $\hat{\Vnl}(\bm{k}) = \int e^{i\bm{k}\cdot\bm{r}}\Vnl(\bm{r})
\mathrm{d}^3r$ is the Fourier transform of the normalized interaction potential
$\hat{\Vnl}(k=0)=1$. The wave number associated with the roton minimum is
denoted as $k_{\mathrm{rot}}$ and $V_1 \leq 0$ and $V_2 \leq 0$ are two
dimensionless parameters that are set to reproduce the dispersion relation of
superfluid $^4$He (see~\cite{Muller2020a}).
This model also includes a beyond mean field correction controlled by two dimensionless parameters $\chi$ and $\gamma$ that correspond to its amplitude and order, respectively. This term arises from considering a strong interaction between bosons.

The parameters used in the simulations were set to $k_{\mathrm{rot}}\xi=1.638$, $V_1 = 4.54$, $V_2=0.01$, $\chi=0.1$ and
$\gamma=2.8$ in order to mimic the dispersion relation of superfluid $^4$He. The speed of sound and the particles density are fixed as $c=1$ and $n_0=1$.

\section{Navier--Stokes simulations}%
\label{app:navier_stokes}

Classical turbulence simulations are performed using the LaTu solver~\cite{Homann2009}, which solves the incompressible Navier--Stokes equations
\begin{align}
  \pdv{\vvec}{t} + \vvec \cdot \gradient \vvec &=
  -\gradient p + \nu \laplacian \vvec + \fvec \\
  \gradient \cdot \vvec &= 0
\end{align}
using a standard Fourier pseudo-spectral method in a three-dimensional periodic domain of size $(2\pi)^3$,
with a third-order Runge--Kutta scheme for the temporal discretization.
Here, $\nu$ is the fluid viscosity, $p$ is the pressure field, and $\fvec$ is an external forcing which emulates a large scale energy injection mechanism.
The forcing is active within a spherical shell of radius $|\kvec| \le 2$ in Fourier space.

Simulations are performed in a grid of $N^3 = 1024^3$ collocation points, at a Taylor scale Reynolds number $\ReLambda \approx 320$.
Circulation statistics are gathered once the simulation reaches a statistically steady state, when the energy injection and dissipation rates are in equilibrium.
Circulation is computed from a set of velocity fields obtained from the simulations.
As in the quantum turbulence simulations, circulation is computed from its velocity-based definition Eq.~\eqref{eq:circulation_intro}, using the Fourier coefficients of the velocity field as described in Appendix~\ref{app:circulation_computation}.

\bibliography{Circulation}

\end{document}